\documentclass[prb,reprint,superscriptaddress,aps,floatfix]{revtex4-2}
\usepackage{amsmath}
\usepackage{amsfonts}
\usepackage{graphicx}
\usepackage{physics}
\usepackage[usenames]{color}
\usepackage{braket}
\usepackage{bbold}
\usepackage{multirow}
\usepackage{tabularx}
\usepackage{array}
\usepackage{hyperref}
\begin{document}

\date{\today}

\title{Discrete Time Crystals in the spin-$s$ Central Spin Model}

\author{Hillol Biswas}
\author{Sayan Choudhury}
\email{sayanchoudhury@hri.res.in}
\affiliation{Harish-Chandra Research Institute, a CI of Homi Bhabha National Institute, Chhatnag Road, Jhunsi, Allahabad 211019}

\begin{abstract}
We propose periodic driving protocols to realize discrete time crystals (DTCs) in a spin-$s$ central spin model. Interestingly, we identify parameter regimes, where eternal period-doubling and higher-order(HO)-DTCs can be realized, even for finite-sized systems. We have determined the dependence of the DTC order on the number of satellite spins and the central spin value, $s$. Intriguingly, we find that certain classes of HO-DTCs produce a series of maximally entangled Bell cat and super-cat states during their dynamical evolution. Finally, we demonstrate that the HO-DTCs can be employed for quantum-enhanced multiparameter sensing at the Heisenberg limit.
\end{abstract}

\maketitle

\section{Introduction}
Periodically driven quantum systems, can exhibit intrinsically non-equilibrium phases of matter that lack equilibrium analogs \cite{bukov2015universal,harper2020topology,rudner2020band,oka2019floquet,weitenberg2021tailoring,banerjee2024emergent}. A paradigmatic example of such a phase is the discrete time crystal (DTC) \cite{khemani2016phase,von2016absolute,yao2017discrete,else2016floquet,sacha2015modeling}, characterized by the spontaneous discrete time-translation-symmetry-breaking (TTSB) and the emergence of subharmonic responses in observables under periodic driving \cite{else2020discrete,zaletel2023colloquium,sacha2017time}. In addition to their foundational significance in the study of non-equilibrium quantum matter, DTCs hold promise for applications in emerging quantum technologies \cite{estarellas2020simulating,moon2024discrete,lyu2020eternal}. Experimental realizations of DTCs have been reported across a range of platforms, including trapped ion arrays \cite{zhang2017observation,kyprianidis2021observation}, Carbon-13 nuclear spins in diamond \cite{choi2017observation,randall2021many,beatrez2023critical}, Rydberg atom systems \cite{bluvstein2021controlling,liu2024higher}, and superconducting qubit processors \cite{mi2022time,bao2024creating,xu2021realizing,frey2022realization}.

The presence of interactions is essential for stabilizing TTSB and sustaining a non-trivial DTC phase~\cite{khemani2019brief,sacha2020time}. However, a major challenge arises from the generic tendency of interacting Floquet systems to undergo unbounded energy absorption from the drive, eventually thermalizing to a featureless infinite-temperature state \cite{d2014long,lazarides2014equilibrium,choudhury2014stability,zhang2015thermalization}. Early realizations of DTCs circumvented this issue by leveraging many-body localization (MBL), which inhibits thermalization and enables the persistence of non-equilibrium order \cite{zhang2017observation}. Subsequently, it has been demonstrated that disorder is not a strict requirement for realizing DTC phases \cite{rovny2018observation,rovny2018p,pal2018temporal,huang2018clean,pizzi2019period,pizzi2021higher,choudhury2021route,munoz2022floquet,russomanno2017floquet,surace2019floquet,maskara2021discrete,sarkar2024time,liu2023discrete,yousefjani2024discrete}. Notably, the Floquet central spin model has emerged as a prominent example of a disorder-free DTC, where long-lived period-doubled oscillations of the central spin magnetization have been both theoretically predicted~\cite{frantzeskakis2023time,kumar2024hilbert,cabot2022metastable} and experimentally confirmed~\cite{pal2018temporal,geng2021ancilla}. In our previous work~\cite{biswas2025floquetcentralspinmodel}, we have explored some of the features of a central spin model when the central spin is a spin-$1/2$ object.

In this work, we explore the dynamics of the driven central spin system for a general spin-$s$ central spin, and classify the various kinds of DTCs that can arise in this system. We propose ways to drive the whole system or a sub-system to the eternal period-doubling DTC phase by choosing the number of satellite spins and the value of the central spin. We also devise a protocol to observe eternal higher-order (HO)-DTC phase where the physical observables have a period more than $2T$. In certain situations within these phases, we find that the system generates highly entangled states such as the GHZ state during their dynamical evolution. Finally, we show that these HO-DTC phases can have quantum-enhanced multiparameter sensing abilities.\\

This paper is organized as follows. We introduce the model and discuss a many-body echo protocol to realize a period-doubling DTC in sec.~\ref{sec:Model}. We then propose and characterize protocols to realize HO-DTCs in sec.~\ref{sec:HODTC}. We demonstrate that these HO-DTCs can be a powerful resource for multi-parameter singing in sec.~\ref{sec:sensor}. We conclude with a summary of our work and a brief discussion of future research directions in sec.~\ref{sec:Conclusion}.

\begin{figure*}
    \centering
    \includegraphics[width=0.9\textwidth]{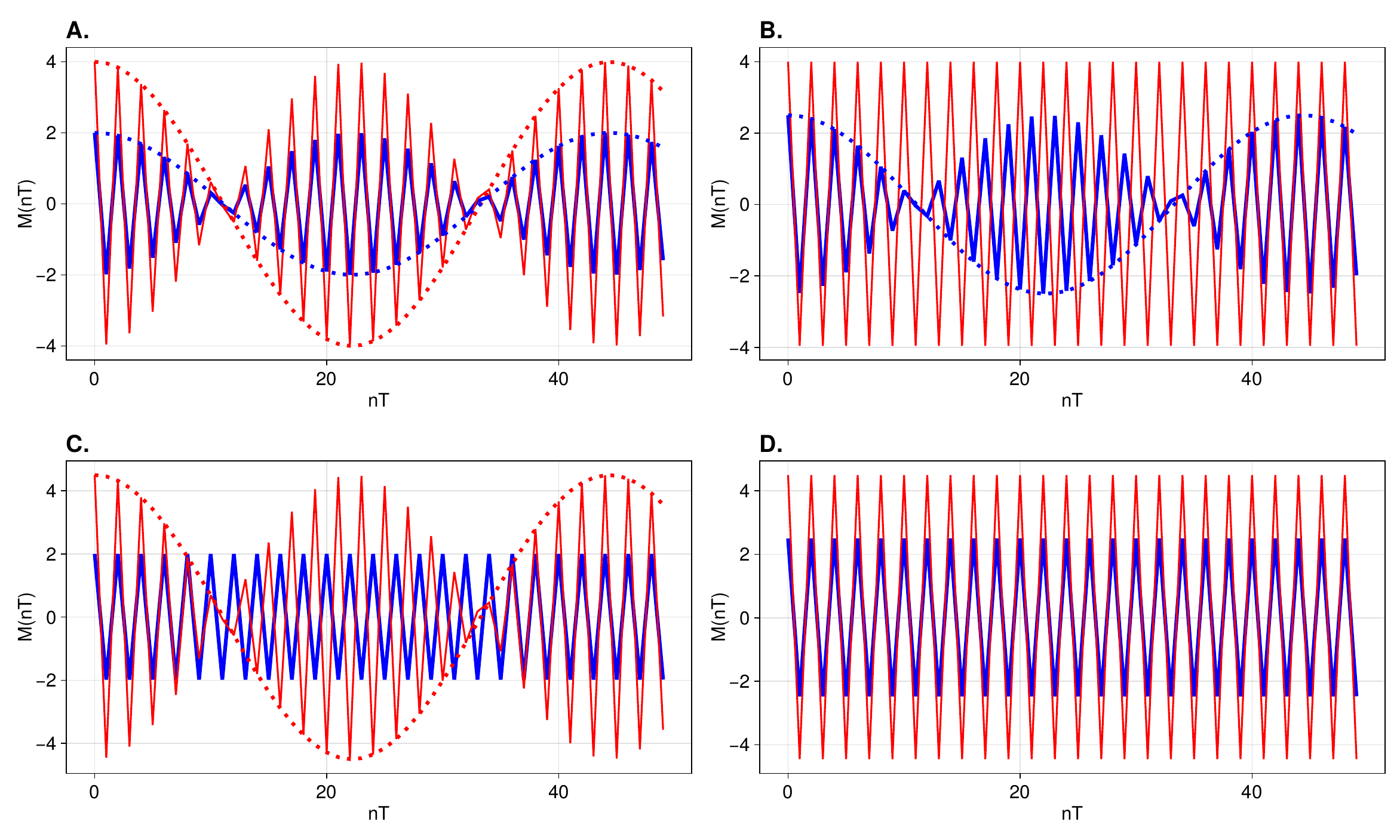}
\caption{\textbf{Magnetization evolution with time of a central spin system at $\lambda=2\pi,~g=3.0$ shows eternal period doubling DTC for finite systems.} (A) $N_{\rm sat}=8;~s=2$, B) $N_{\rm sat}=8;~s=5/2$, C) $N_{\rm sat}=9;~s=2$, D) $N_{\rm sat}=9;~s=5/2$). The blue line denotes central spin magnetization and the red line denotes satellite spin magnetization. The dotted lines follow $M(2nT)$ to exhibit the sinusoidal variation of magnetization at $2nT$.}
\label{M_variation}
\end{figure*}
\begin{figure*}
    \centering
    \includegraphics[width=0.77\textwidth]{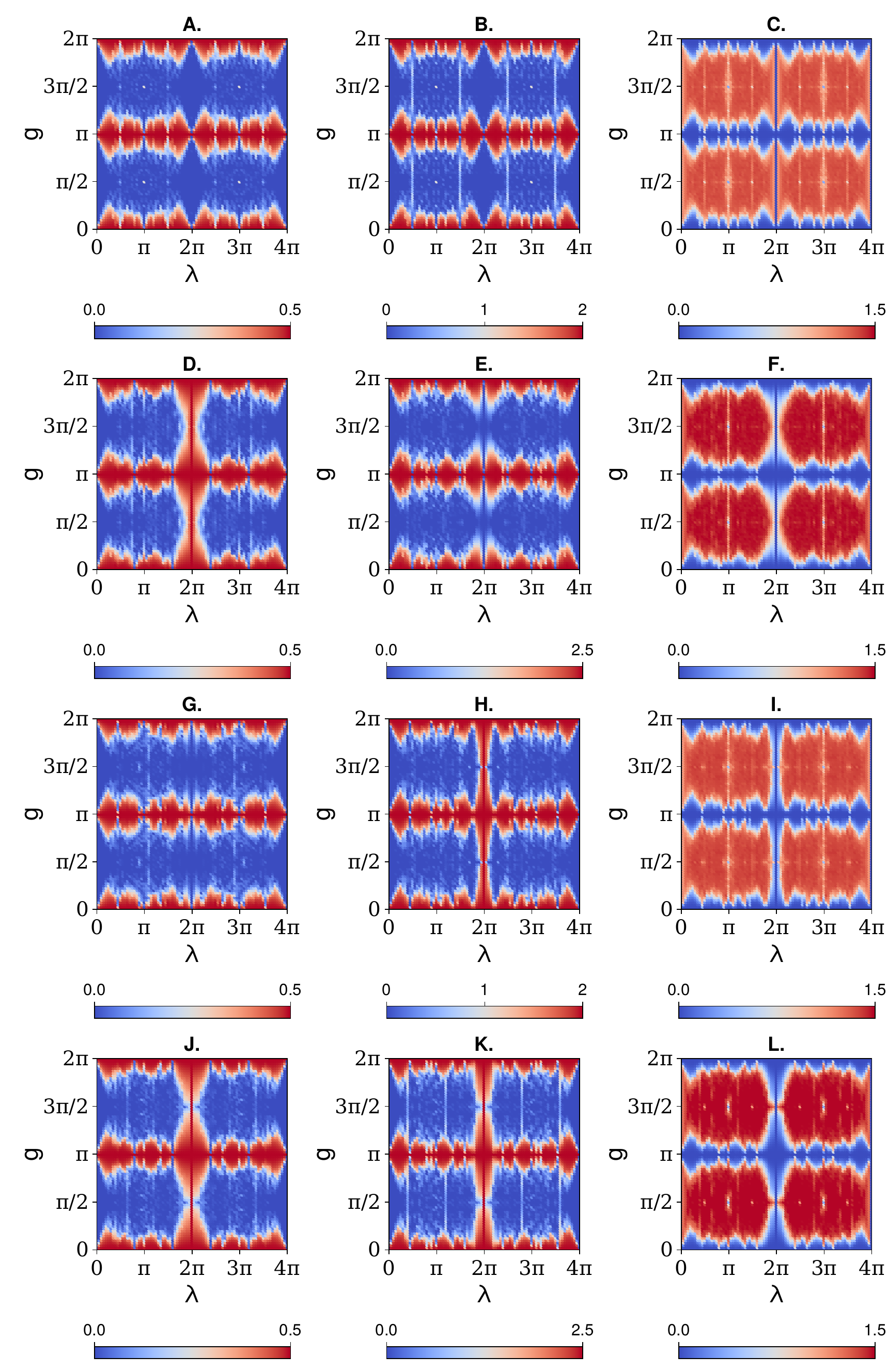}
\caption{\textbf{Regions of non-ergodic behaviour in the $\lambda$ vs $\pi$ phase diagram:} The left-hand panels (A,D,G,J) are for satellite spin magnetization, the central panels (B,E,H,K) are for central spin magnetization and the right-hand panels (C,F,I,L) are for entanglement entropy. From the top to the bottom, the rows represent $N_{\rm sat}=8,~s=2;~N_{\rm sat}=8,~s=5/2;~N_{\rm sat}=9,~s=2;~N_{\rm sat}=9,~s=5/2;~$}
\label{op}
\end{figure*}

\begin{figure*}
    \centering
    \includegraphics[width=0.9\textwidth]{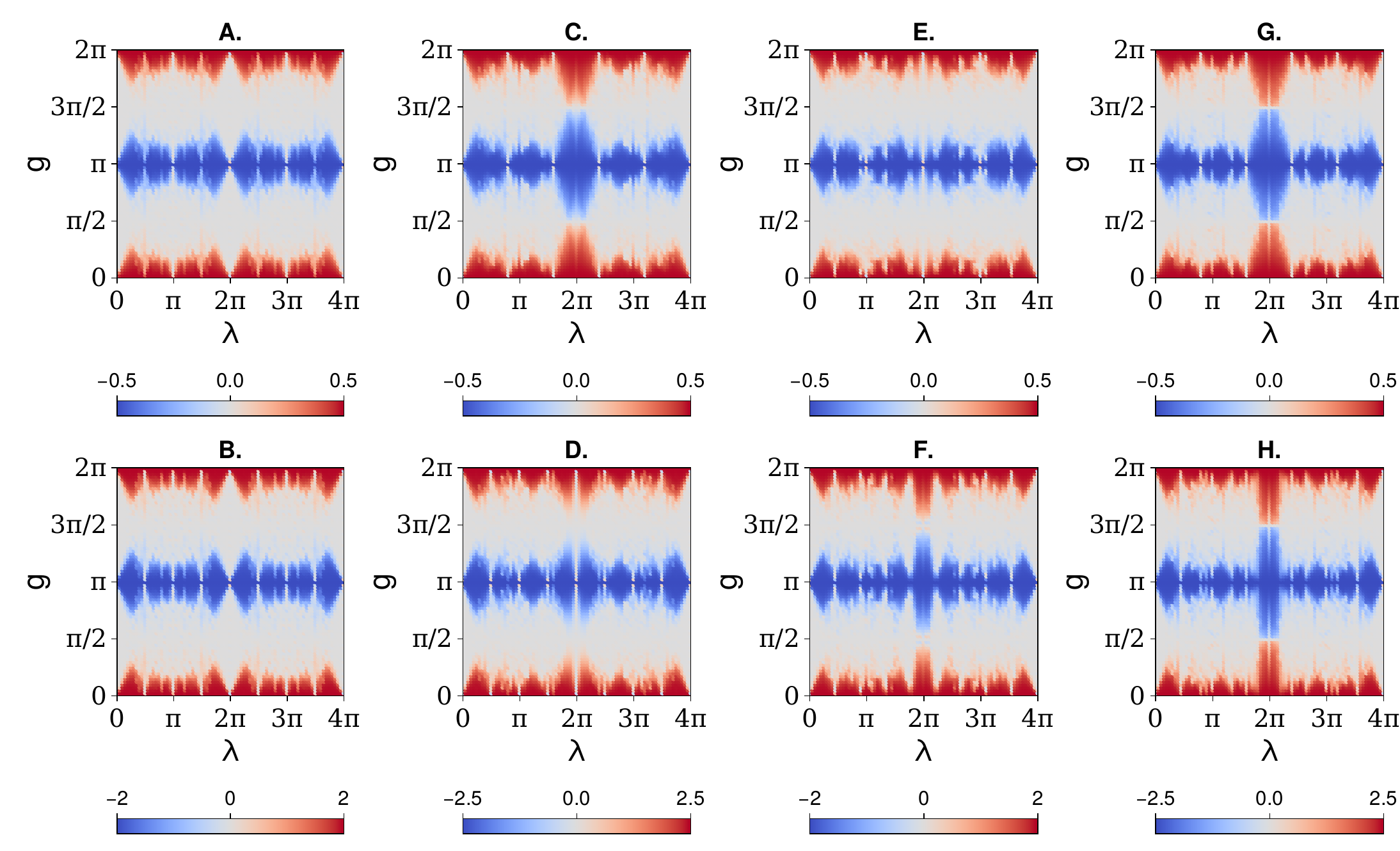}
\caption{\textbf{Two kinds of non-ergodic phenomena are characterized using the relative order parameter.}  The difference between DTC order parameter and DMF order parameter as a function of $\lambda$ and $g$. The left-hand panels (A, C, E, G) are for satellite spin magnetization and the right hand panels (B, D, F, H) are for central spin magnetization. From the top to the bottom, the rows represent $N_{\rm sat}=8,~s=2;~N_{\rm sat}=8,~s=5/2;~N_{\rm sat}=9,~s=2;~N_{\rm sat}=9,~s=5/2;~$}
\label{afm-fm}
\end{figure*}
\begin{figure*}
    \centering
    \includegraphics[width=0.9\textwidth]{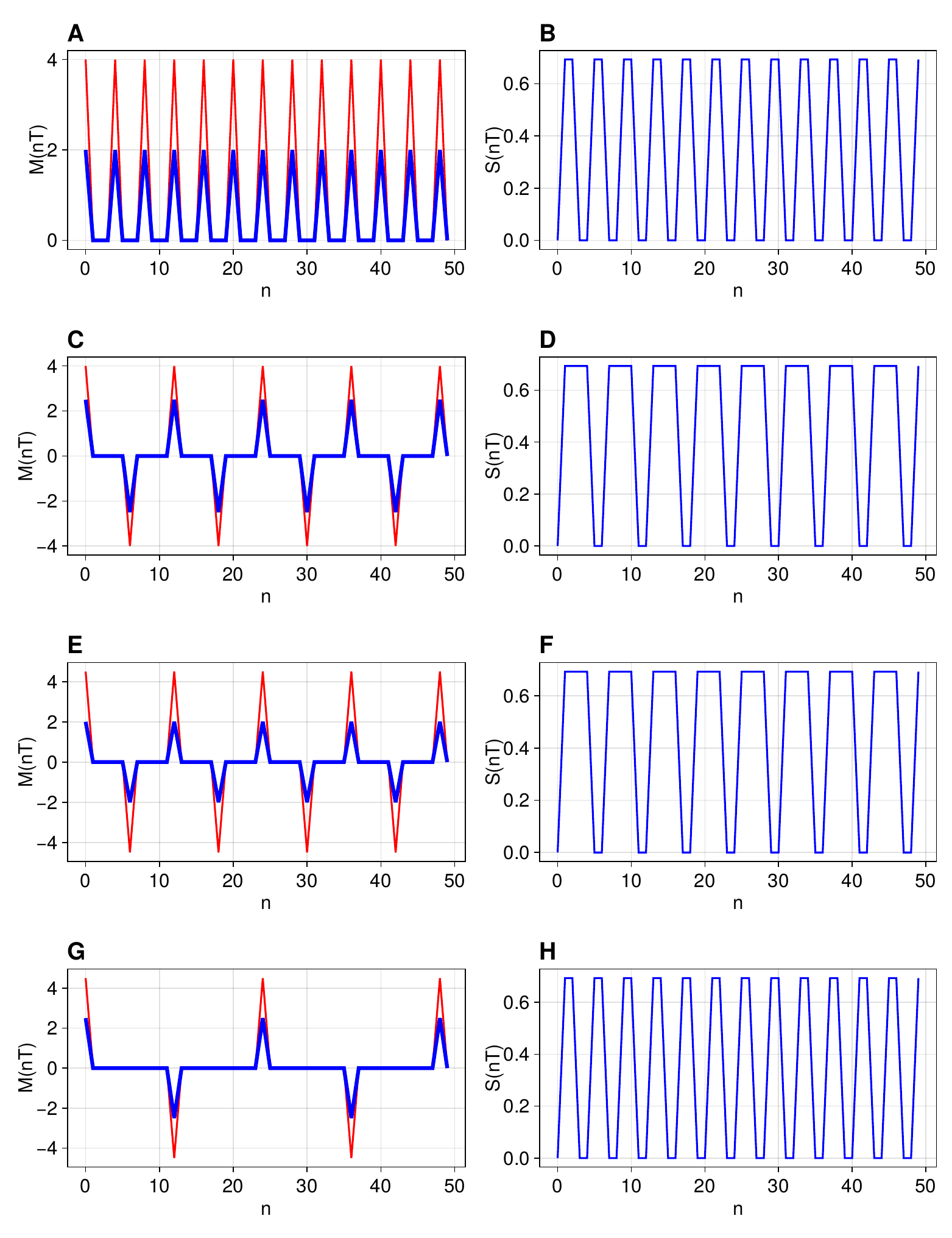}
\caption{\textbf{Magnetization evolution with time for $\lambda=\pi,~g=\pi/2$ (special HO-DTC). The true non-equilibrium nature of this phase can be seen in the entanglement entropy variations.}  top row: $N_{\rm sat}=8;~s=2$, Second row: $N_{\rm sat}=8;~s=5/2$, Third row: $N_{\rm sat}=9;~s=2$, Fourth row: $N_{\rm sat}=9;~s=5/2$). The left-hand panel shows the magnetization and the right hand panel shows the entanglement entropy variation. The blue line denotes central spin magnetization, and the red line denotes satellite spin magnetization.}
\label{ho-DTC}
\end{figure*}
\section{Model and Period-doubling DTC}
\label{sec:Model}
We study a periodically driven spin-$s$ central spin model:
\begin{equation}
H(t) =
     \begin{cases}
			H_d= & \frac{2}{T}{} (g_s\sum_{i=1}^{N_{\rm sat}} S_i^z+g_c S_c^z) {\rm~for~} t\in[0,~T/2)\\
			H_0=& -2\lambda \sum_{i=1}^{N_{\rm sat}} S_i^x S_c^x {\rm~for~} t\in[~T/2,~T),
      \end{cases}
    \label{eq:Hamiltonian}
\end{equation}
where $S_i^{\alpha} = \frac{1}{2} \sigma_i^{\alpha}$ represent the spin-1/2 satellite spins and $S_c$ is the spin-$S$ central spin and it is represented by the $(2s+1) \times (2s+1)$ Pauli spin-$s$ operators.\\

The unitary evolution operators associated with the Hamiltonian are:
\begin{align}
    &U_0=\exp(-iH_0T)=\exp(i\lambda\sum_{i=1}^{N_{\rm sat}}S^x_iS_c^x)\\ \nonumber
    &U_d=\exp(-iH_d)=\exp(-ig_c S_c^z)\Pi_{i=1}^{N_{\rm sat}}\exp(-ig_s S^z_i) 
\end{align}
For convenience, we write $U_d = U_c U_{\rm sat}$ into two parts: $U_c=\exp(-ig_c S_c^z)$ and $U_{\rm sat}=\Pi_{i=1}^{N_{\rm sat}}\exp(-ig_s S^z_i)$, where $U_c$ and $U_{\rm sat}$ commute.\\

Before proceeding further, it is useful to recall the salient characteristics of a Discrete time crystal. A discrete time crystal is traditionally characterized by three features: \\
\begin{enumerate}
\item Discrete TTSB: the system should exhibit discrete TTSB and physical observables should oscillate with a period $nT$ where $n>1$ for generic initial states, without fine-tuning the Hamiltonian parameters.
\item Rigidity: The oscillations must remain in phase up to arbitrarily long times in the thermodynamic limit.
\item Crypto-equilibrium: the system is in equilibrium in a time-dependent frame, leading to no growth of the entropy at long times.
\end{enumerate}
In this section, we will demonstrate the existence of period-doubling DTCs, which satisfy all of these criteria are fulfilled. In the next section, we will demonstrate the existence of HO-DTCs that go beyond the paradigm of crypto-equilbrium.\\

We now analyze the dynamics of this system when $\lambda = 2 \pi$. In this case, the system can exhibit eternal period-doubling dynamics when the central spin value, $s=1/2$. In order to analyze the general spin-$s$ model, we first start by looking at the time evolution operator over two drive periods:
\begin{align}
    U^2=U_0U_dU_0U_d=U_0U_cU_{\rm sat}U_0U_{\rm sat}U_c.
\end{align}
We now note that when $N_{\rm sat}$ is even (odd) $U_{\rm c}$ commutes (anti-commutes) with $U_0$. And, if the dimension of $S_c$ is even (odd), i.e. the central spin is a full (half) integer, $U_{\rm sat}$ commutes (anti-commutes) with $U_0$. Hence:\\
\begin{widetext}
\begin{align}
   \label{cases}
    U^2=\begin{cases}
        U^2_0~\mathbb{1}_{\rm sat}\otimes \mathbb{1}_{\rm c}{\rm~~for~odd~}N_{\rm sat}{\rm~and~half~integer~}s\\
        U^2_0~\Pi_{i=1}^{N_{\rm sat}}\exp(-i2g_s S^z_i)\otimes \mathbb{1}_{\rm c}{\rm~~for~odd~}N_{\rm sat}{\rm~and~full~integer~}s\\
        U^2_0~\mathbb{1}_{\rm sat}\otimes \Pi_{i=1}^{N_{\rm c}}\exp(-i2g_c S_c^z){\rm~~for~even~}N_{\rm sat}{\rm~and~half~integer~}s\\
        U^2_0~\Pi_{i=1}^{N_{\rm sat}}\exp(-i2g_s S^z_i)\otimes \Pi_{i=1}^{N_{\rm c}}\exp(-i2g_c S_c^z){\rm~~for~even~}N_{\rm sat}{\rm~and~full~integer~}s\\
    \end{cases}
\end{align}
\end{widetext}
For an odd number of satellite spins and a half-integer value of the central spin, $U^2_0$ becomes an identity matrix. Hence, we will get a perfect revival of the state of the complete system, and both the central spin and the satellite spins will show an exact period doubling DTC phase. For an odd number of satellite spins and a full-integer value of the central spin the satellite spins will show sinusoidal oscillation while the central spin will be in the exact period doubling DTC phase. For an even number of satellite spins and a half-integer value of the central spin the satellite spins will be in the exact period doubling DTC phase while the central spin will show sinusoidal evolution. Finally, when the system has an even number of satellite spins and the central spin has a full-integer value, both will show sinusoidal evolution. The magnetization plots in Fig.~\ref{M_variation} and table \ref{tab_period-doubling} summarizes the period doubling phase for various values of central spin and satellite spins.\\
\begin{table}[]
    \centering
    \begin{tabular}{|m{7mm}|m{7mm}|m{20mm}|m{20mm}|m{20mm}|}
\hline 
    $N_{\rm sat}$ & $s$ & $M_{\rm sat}$ variation & $M_{\rm c}$ variation & Type of DTC \\
    \hline\hline
    even & even & sinusoidal & sinusoidal & Rabi oscillation \\ \hline
    even & odd & period doubling & sinusoidal & sub-system DTC \\ \hline
    odd & even & sinusoidal & period doubling & sub-system DTC \\ \hline
    odd & odd & period doubling & period doubling & eternal DTC \\ \hline
    \end{tabular}
    \caption{Types of period-doubling DTC at $\lambda=2\pi$ for various values of $N_{\rm sat}$ and $s$}
    \label{tab_period-doubling}
\end{table}

This complete revival or sinusoidal variation can be seen exactly at $\lambda=2\pi$. To see what happens away from this point, we start with fully ${\rm x}$-polarized initial state and we plot $2nT:  \langle \overline{M} \rangle =\frac{1}{N} \sum_{n=1}^{N} M(2nT)$, (where $M$ is average magnetization) for the satellite spins (Fig.~\ref{op} A, D, G, J) and the central spin (Fig.~\ref{op} B, E, H, k). We have also plotted the entanglement entropy of the system in Fig.~\ref{op} C, F, I, L.\\

In Fig.~\ref{op} D, H, J and K, we can see that the average magnetization is high at $\lambda=2\pi$ because of the period doubling oscillations, the $N_{\rm sat}$ and $s$ values are consistent with table \ref{tab_period-doubling}. In Fig.~\ref{op} A, B, E and G, $\lambda=2\pi$ region shows very small average magnetization, which is consistent with the expected sinusoidal variation. But for all values of $N_{\rm sat}$ and $s$, the central spin and the satellite spins remain disentangled at $\lambda=2\pi$, as expected from eq \ref{cases}. It corresponds to the vertical blue region at $\lambda=2\pi$ in the entanglement entropy plots.\\

The entanglement entropy plots also show that the system becomes disentangled at $g=0/2\pi, \pi$. The absence of entanglement between the central spin and the satellite spins can be attributed to two different non-ergodic phenomena: a) dynamical many-body freezing (DMF) when the system approximately returns to its initial state, and b) a DTC when discrete time translation symmetry is broken. These two phases can be characterized (Fig.~\ref{afm-fm}) by a new Relative order parameter:
\begin{eqnarray}
    O (nT) &=& O_{\rm DTC}(nT)- O_{\rm DMF} (nT), \,\,\, {\rm where} \nonumber\\
    O_{\rm DTC} (nT) &=& (-1)^n M(nT)\,\,  {\rm and}\, \, O_{\rm DMF} (nT) = M(nT).\nonumber \\
    \label{eq:DTC-DMF}
\end{eqnarray}
When the relative order parameter $O\sim -0.5(-s)$ for the satellite spins (central spin) the system shows DTC phase, when $O\sim0.5(s)$ for the satellite spins (central spin) the system is in DMF regime.\\

\section{Higher order DTC}
\label{sec:HODTC}

In the entanglement entropy plots in Fig.~\ref{op} (C, F, I, L), other than the DTC and the DMF phase, 4 symmetrically placed small regions of low entanglement can be seen at $(\lambda ,g)=(\pi,\pi/2),(3\pi,\pi/2),(\pi,3\pi/2)~ \rm and~(3\pi,3\pi/2)$. At these points, we can see higher-order DTC phases with time period $nT$ where $n>2$. When $N_{\rm sat}$ is odd and $s$ is a half-integer, we get period 24 DTC. We get period 12 DTC when $N_{\rm sat}$ is odd (even) and $s$ is a full-integer (half-integer). When $N_{sat}$ is even and $s$ is a full integer, we get a new kind of DTC phase that has a period of $4T$. The speciality of this new phase is that the polarity of magnetization is never reversed. The magnetization variation as a function of time at $\lambda=\pi,~g=\pi/2$ is plotted in Fig.~\ref{ho-DTC} (A, C, E, G). The time periods for various values of $N_{\rm sat}$ and $s$ are summarized in table \ref{tab_ho-dtc}.\\

\begin{table}[]
    \centering
    \begin{tabular}{|c|c|c|}\hline
         $N_{\rm sat}$ & $s$ & DTC period  \\ \hline\hline
        even & integer & 4\\
        even & half integer & 12\\
        odd & integer & 12\\
        odd & half integer & 24\\ \hline
    \end{tabular}
    \caption{Varying time period of the special higher order DTC depending on $N_{\rm sat}$ and $s$}
    \label{tab_ho-dtc}
\end{table}

We try to understand the nature of this phase by plotting the magnetization and entanglement entropy as a function of time (Fig.~\ref{ho-DTC} B, D, F, H). It shows that the central and the satellite spins become entangled as we start driving the system, and after a certain number of periods, they become disentangled again. The genuine growth and decay of the entanglement in the system indicates that this DTC is beyond the regime of crypto-equilibrium and represents a truly non-equilibrium phase of matter.\\

We proceed to study the specific states the system goes through as we start driving it. For that, we start with a fully x-polarized state: $\vert \psi(t=0) \rangle=\left(\prod_{i=1}^{N_{\rm sat}}\vert {\rm +x} \rangle_s^i\right)\otimes \vert s \rangle^x_c$. At $g=\pi/2$ the evolution operator for each of the satellite spins will become $\exp(-i\dfrac{\pi}{2}S_i^z)$ and it will take the spins from $\vert \pm x\rangle$ to $\vert \pm y\rangle$ state and $\vert \pm y\rangle$ to $\vert \mp x\rangle$ state. Then we calculate the dynamic phase acquired by each term due to the satellite spin-central spin interaction.\\

\textit{Case I: When $N_{\rm sat}$ is odd and $s$ is a full integer:}\\

In this case, following the recipe mentioned above, at $t=3T$ we will get an entangled yet symmetric state:
\begin{align}
    \vert \psi(3T) \rangle
    &=\left(\prod_{i=1}^{N_{\rm sat}}\vert {\rm +y} \rangle_s^i\right) \vert+s \rangle_c^x-\left(\prod_{i=1}^{N_{\rm sat}}\vert {\rm +y} \rangle_s^i\right) \vert-s \rangle_c^x\\ \nonumber &-i\left(\prod_{i=1}^{N_{\rm sat}}\vert {\rm -y} \rangle_s^i\right) \vert-s \rangle_c^x-i\left(\prod_{i=1}^{N_{\rm sat}}\vert {\rm -y} \rangle_s^i\right) \vert-s \rangle_c^x
\end{align}
Clearly, this state has equal probability that the satellite spin can be polarized along the +y or -y direction and the central spin can also be polarized along the +x or -x direction with equal probability. Hence the magnetization will be $0$. we call these kind of states, Bell super-cat states.\\

Following the same procedure, we find the system will completely reverse its initial polarization at $t=6T$ and we get a product state where all the spins polarized along the -x direction, marking the completion of half-period. From here, it is evident that driving the system $6$ more times will again reversed its polarization and the system will return to its original +x polarized state, thus exhibiting a 12-DTC phase. This is why at $6T$ and $12T$ the systems shows minimal entanglement entropy (Fig.~\ref{ho-DTC}C,D).\\

\textit{Case II: When $N_{\rm sat}$ is even and $s$ is a half-integer:}\\

This is quite similar to the previous case except that the alignment of the satellite spins and the central spin will get interchanged. For example, at $t=3T$ we will get
\begin{align}
    \vert \psi(3T) \rangle
    &=\left(\prod_{i=1}^{N_{\rm sat}}\vert {\rm +x} \rangle_s^i\right) \vert+s \rangle_c^y-\left(\prod_{i=1}^{N_{\rm sat}}\vert {\rm +x} \rangle_s^i\right) \vert-s \rangle_c^y\\ \nonumber &-i\left(\prod_{i=1}^{N_{\rm sat}}\vert {\rm -x} \rangle_s^i\right) \vert-s \rangle_c^y-i\left(\prod_{i=1}^{N_{\rm sat}}\vert {\rm -x} \rangle_s^i\right) \vert-s \rangle_c^y
\end{align}
The rest of the features will be exactly similar to the previous case: the system will reverse its polarization at $t-6T$ and return to its initial state at $t=12T$, showing a 12-DTC phase. It also has entanglement entropy $=0$ at $6T$ and $12T$ (Fig.~\ref{ho-DTC}C,D).\\

\textit{Case III: When $N_{\rm sat}$ is odd and $s$ is a half-integer:}\\

The evolution is a little more complicated in this case because the system goes through a wide of states and exhibits a 24-DTC phase. Just like the previous two cases, this time also, the system goes through a set of maximally entangled Bell super-cat states at $t=T,2T,3T$ etc. But, in this case, at quarter period, i.e. at $t=6T$ we will get a GHZ state of the entangled system:
\begin{align}
    \label{psi_6T}
    \vert \psi(6T)\rangle&=(-1-i)\left(\prod_{i=1}^{N_{\rm sat}}\vert {\rm +x} \rangle_s^i\right)\vert +s \rangle_c^x\\ \nonumber &+(1-i)\left(\prod_{i=1}^{N_{\rm sat}}\vert {\rm -x} \rangle_s^i\right) \vert -s \rangle_c^x
\end{align}

Half-period is again marked with an expected reverse polarized state at $t=12T$. And the system returns to its initial state at $t=24T$. Similar to the previous cases we see the entanglement entropy vanish at these times: $12T$ and $24T$. In addition to these moments, at $t=4nT$ where $n=1,2,4,5$ the entanglement vanishes in Fig.~\ref{ho-DTC}H. At these times, the system arrives at a product of 2 separate GHZ states associated with the central spin and the satellite spins. For example:
\begin{align}
    \vert \psi(4T)\rangle&=\Bigg[\left(\prod_{i=1}^{N_{\rm sat}}\vert {\rm -z} \rangle_s^i\right)-\left(\prod_{i=1}^{N_{\rm sat}}\vert {\rm +z} \rangle_s^i\right)\Bigg]\\ \nonumber &\otimes\Bigg[\vert +s \rangle_c^z-\vert -s \rangle_c^z\Bigg]
\end{align}

\begin{figure}
    \centering
    \includegraphics[width=0.9\linewidth]{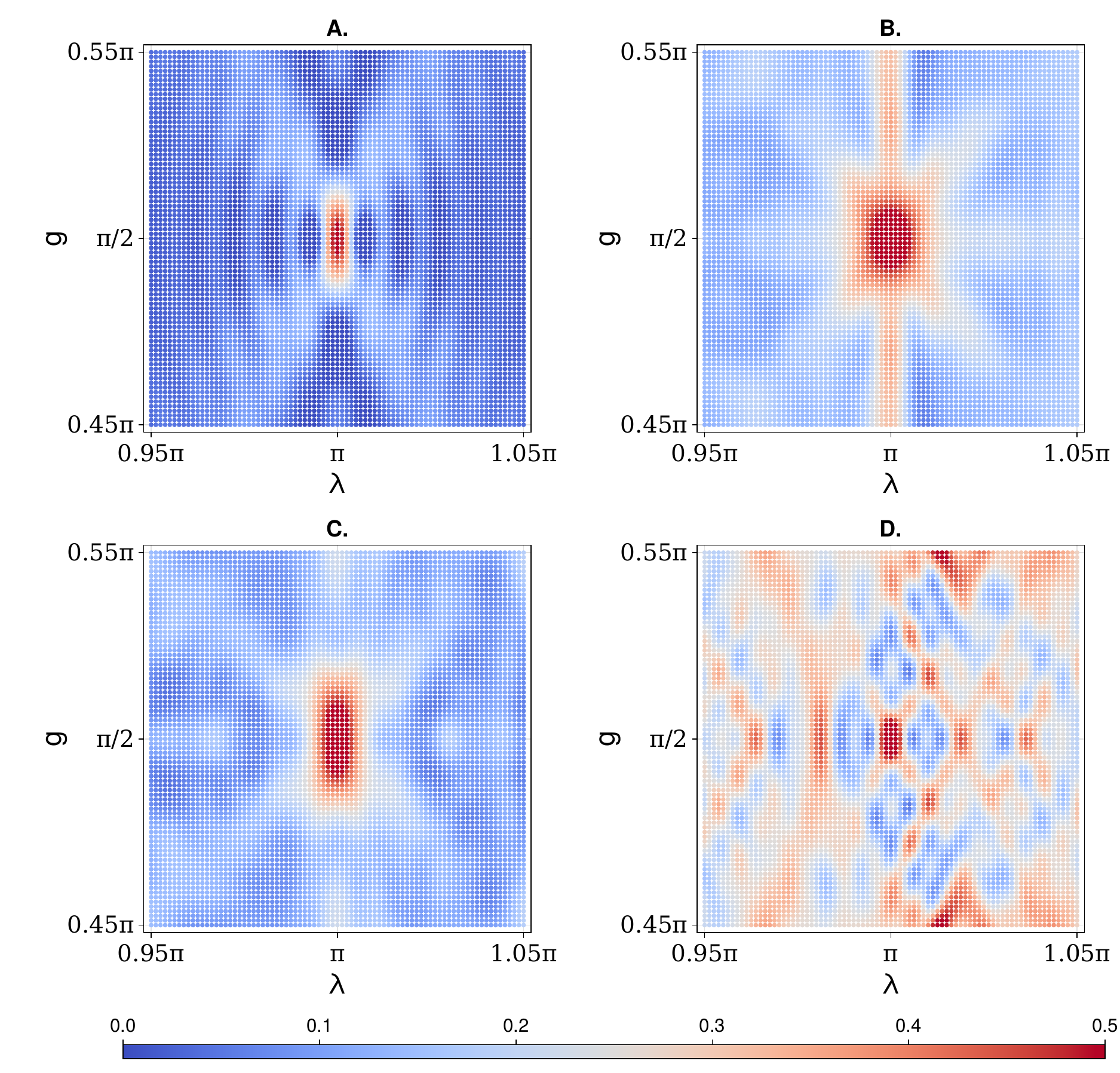}
    \caption{\textbf{Order parameter for special HO-DTC shows the narrow window in the parameter space where this phase exists.} A. even $N_{\rm sat}$ and full integer $s$, B. even $N_{\rm sat}$ and half integer $s$, C. odd $N_{\rm sat}$ and full integer $s$, D. odd $N_{\rm sat}$ and half integer $s$}
    \label{ho-op}
\end{figure}

\textit{Case IV: When $N_{\rm sat}$ is even and $s$ is a full integer:}\\

This is a special case since there is no reversal of polarization. In this case the system goes through two Bell super-cat states:
\begin{align}
    \psi(T)&=\left(\prod_{i=1}^{N_{\rm sat}}\vert {\rm +y} \rangle_s^i\right)\vert+s\rangle_c^y+\left(\prod_{i=1}^{N_{\rm sat}}\vert {\rm -y} \rangle_s^i\right)\vert +s \rangle_c^y\\ \nonumber &+\left(\prod_{i=1}^{N_{\rm sat}}\vert {\rm +y} \rangle_s^i\right)\vert -s \rangle_c^y-\left(\prod_{i=1}^{N_{\rm sat}}\vert {\rm -y} \rangle_s^i\right)\vert -s \rangle_c^y
\end{align}
\begin{align}
        |\psi(2T)\rangle=&\left(\prod_{i=1}^{N_{\rm sat}}\vert {\rm +x} \rangle^s_i\right) \vert +s \rangle^x_c-\left(\prod_{i=1}^{N_{\rm sat}}\vert {\rm +x} \rangle^s_i\right) \vert -s \rangle^x_s\\ \nonumber &-\left(\prod_{i=1}^{N_{\rm sat}}\vert {\rm -x} \rangle^s_i\right) \vert +s\rangle^x_s+\left(\prod_{i=1}^{N_{\rm sat}}\vert {\rm -x} \rangle^s_i\right) \vert -s \rangle^x_c
\end{align}
After the third period, the system polarizes in the y direction: $\psi(3T)=\left(\prod_{i=1}^{N_{\rm sat}}\vert {\rm -y} \rangle^s_i\right) \vert { -s} \rangle^y_c$. The fourth kick will take $|-y\rangle$ states to $\vert +x\rangle$. Hence, after 4 periods, the system will return to its initial x-polarized disentangled state, thus exhibiting a 4-period DTC phase. This is why in Fig.~\ref{ho-DTC}A and B entanglement entropy vanishes at every $(4n-1)T$ and $4nT$.\\

This is a very subtle phase and can be detected only within a very narrow window in the parameter regime, as seen in Fig. \ref{ho-op}, where we plot an order parameter $\langle \overline{M} \rangle =\frac{1}{N} \sum_{n=1}^{N} M(\alpha nT)$, $\alpha=4,~12~\rm or~24$ consistent with table \ref{tab_ho-dtc}. \\

Since the system goes through a set of highly symmetric states, we call this phase the special HO-DTC phase. A detailed analytical description of this phase is given in the supplemental material.\\

There are two more subtle classes of higher-order DTC phases only for certain values of $N_{\rm sat}$ and $s$ as mentioned in table \ref{tab_ho-dtc2} and \ref{tab_ho-dtc3}. But there is a significant difference between these two higher order DTC phases and the previous one: unlike the previous case, in these two phases, the magnetization does not always go to $0$ between two disentangled and polarized states. In the HO-DTC phase at $\lambda=\pi,~g=\pi/2$, the application of each kick takes the system to maximally entangled Bell cat or super-cat states before it eventually comes back to its initial state. But, the magnetization graph for these two HO-DTC phases, suggests that the states in between two polarized states are not so special. The entanglement entropy graphs supports that. This is why we call them regular HO-DTC phases. More details about these phases are given in the supplemental material.\\

\begin{figure}
    \centering
    \includegraphics[width=0.45\textwidth]{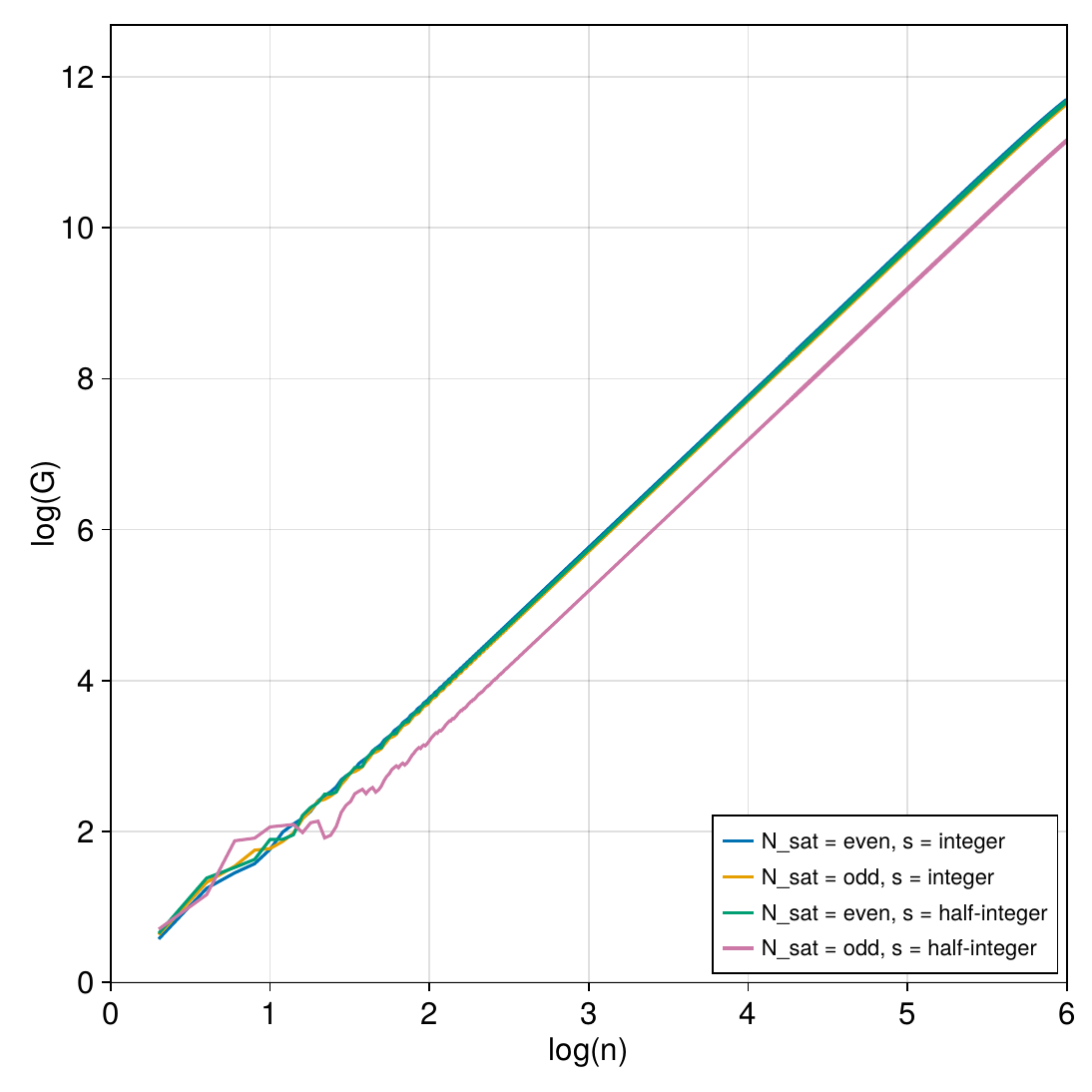}
    \caption{Dynamic growth of G with time for a fixed value of $N_{\rm sat}$ and $s$ shows quadratic increment that indicates quantum enhanced sensing ability.}
    \label{qfi-kicks}
\end{figure}

\begin{figure*}
    \centering
    \includegraphics[width=0.9\textwidth]{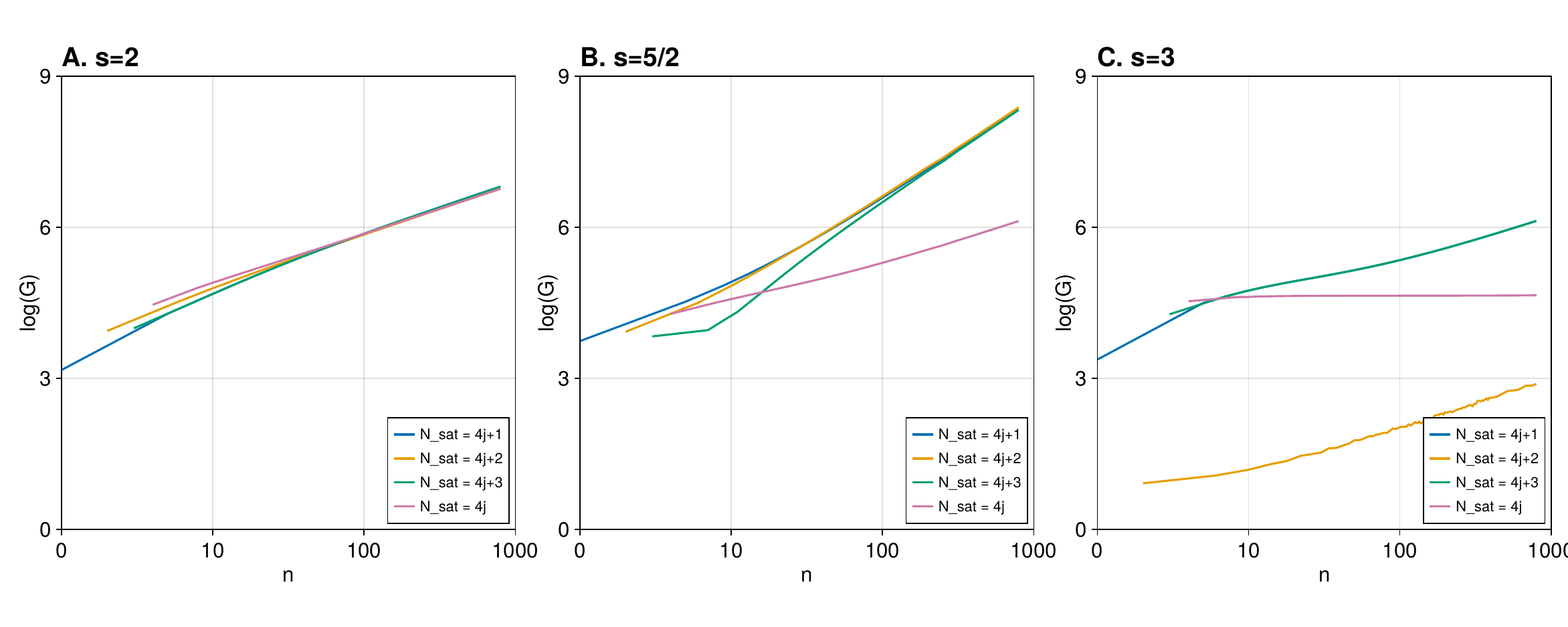}
    \caption{\textbf{Growth of G with value of $N_{\rm sat}$ for a fixed $s$:} Only for half-integer central spin, the quantum enhanced sensing abilities can be achieved.}
    \label{qfi-n}
\end{figure*}

\section{Quantum Multi-parameter sensing}
\label{sec:sensor}

In the previous section we saw that the HO-DTC phase at $\lambda=\pi,~g=\pi/2$ shows a sharp and periodic change in entanglement between the central spin and the satellite spins for a very narrow window in the $\lambda \sim g$ parameter space. This makes the system a perfect candidate for sensing both interaction strength ($\lambda$) and the magnetic field ($g$) simultaneously. In this section we characterise the sensing capabilities using $2\times 2$ Quantum Fisher Information matrix $\mathcal{F}$ whose elements are: 
\begin{widetext}
\begin{align}
    &\mathcal{F}_{\lambda g}= 4~\rm{Re} \bigg[\frac{\langle\psi_{\lambda+\delta \lambda, g}|\psi_{\lambda, g+\delta g}\rangle-\langle\psi_{\lambda+\delta \lambda, g}|\psi_{\lambda, g}\rangle\langle\psi_{\lambda, g+\delta g}|\psi_{\lambda, g}\rangle}{(\delta\lambda\delta g)} \bigg]
\end{align}
\end{widetext}
The equally-weighted uncertainty in measuring $\lambda$ and $g$ is given by $\delta \lambda^2 +\delta g^2 \ge G^{-1}$, where
\begin{equation}
G = \frac{\mathcal{F}_{\lambda\lambda} + \mathcal{F}_{gg}}{\mathcal{F}_{\lambda\lambda}\mathcal{F}_{gg} - \mathcal{F}_{\lambda g}\mathcal{F}_{g \lambda }}.
\label{eq:QFI}
\end{equation}

The scaling behaviour of $G(t=nT,N_{\rm sat})$ can be characterized using two exponents associated with $n$ and $N_{\rm sat}$: $G \propto n^{\alpha}N_{\rm sat}^{\beta}$. For classical sensors, the best possible sensitivity corresponds to $\alpha=\beta=1$, which is known as the standard quantum limit.\\

We have plotted $\ln{G}$ at $\lambda=\pi$ and $g=\pi/2$ as a function of time in Fig.~\ref{qfi-kicks} to show its dynamic growth. It increases quadratically with time: $\alpha=2$\\

In Fig.~\ref{qfi-n} we have plotted the growth of $G$ with increasing number of satellite spins for various values of $s$. We notice that $\beta$ can have 3 different values: $0,~1,~2$ for various values of $s$. When $s$ is an odd integer, $\beta$ is always $1$ except for $N_{\rm sat}=4j$ when $\beta =0$. When $s$ is an even integer, $\beta$ is always $1$. And finally, when $s$ is a half-integer the scaling is the best: $\beta=2$ unless $N_{\rm sat}=4j$ when $\beta=1$.\\

Hence, for certain values of $N_{\rm sat}$ and $s$ the sensitivity in our protocol reaches the Heisenberg limit $\alpha=\beta=2$.\\

\section{Conclusion and Outlook}
\label{sec:Conclusion}

In this work, we present a scheme to obtain time crystals by periodically driving a central spin system. We can choose to drive a certain subsystem or the whole system into period doubling DTC phase by fixing the number of satellite spins and the value of the central spin. This DTC phase lasts up to infinite time for a particular value of the interaction strength ($\lambda=2\pi$), even for finite-sized systems. Furthermore, at $\lambda=(2j+1)\pi$ and $g=(j'+1/2)\pi$ where $j,j^{\prime} \in \mathbb{Z}$), we see eternal higher order DTC phase. The entanglement entropy goes back and forth between the lowest possible value (associated with disentangled central and satellite spins) to the highest possible value (associated with the formation of maximally entangled GHZ and GHZ-like states). Thus, our protocol provides a recipe for a swift preparation of GHZ states. We also find two other regimes of eternal higher-order DTC phase at other places in the $\lambda\sim g$ parameter space, which do not produce GHZ states. In the end we show that the central spin system can be very useful for multiparameter quantum metrology. \\

In the future, it would be interesting to incorporate interaction among the satellite spins and examine whether this destroys the eternal DTC phase or provides more ways to control it.
%\newpage
\bibliographystyle{apsrev4-2}
\bibliography{ref}

\newpage
\clearpage
\begin{widetext}
\renewcommand{\thepage}{S\arabic{page}} 
\renewcommand{\thesection}{S\arabic{section}}  
\renewcommand{\thetable}{S\arabic{table}}  
\renewcommand{\thefigure}{S\arabic{figure}}
\renewcommand{\theequation}{S\arabic{equation}}
\setcounter{figure}{0}
\setcounter{page}{1}
\setcounter{equation}{0}

\begin{center}
\textbf{\Large{Supplemental Material for `Discrete Time Crystals in the spin-$s$ Central Spin Model'}}
\end{center}

In the supplemental material, we provide a detailed analysis of the route to realize the higher-order(HO)-DTC that is useful for multi-parameter sensing and some results on other HO-DTC phases that arise in this system. We note that for all of these HO-DTCs exhibit entanglement oscillations and they go beyond the standard crypto-equilibrium paradigm of other DTCs.\\

\section*{DETAILED ANALYSIS OF THE MECHANISM TO REALIZE A HIGHER-ORDER TIME CRYSTAL}
In this section, we provide an analytical description of this phase. We assume the initial state to be completely polarized in the x direction: $\vert \psi(t=0) \rangle=\left(\prod_{i=1}^{N_{\rm sat}}\vert {\rm +x} \rangle_s^i\right)\otimes \vert s \rangle^x_c$. At $g=\pi/2$, $U_{\rm sat}$ becomes $\Pi_{i=1}^{N_{\rm sat}}\exp{(-i\dfrac{\pi}{2}S_i^z)}$ and it takes each spin $1/2$ particle from $|+x\rangle$ to $|+x\rangle-i|-x\rangle$. $U_{\rm c}=\exp(-i\dfrac{\pi}{2} S_c^z)$ becomes a diagonal matrix whose $l^{th}$ entry is $\exp{(-i\dfrac{(s+1-l)\pi}{2})}$. It will convert fully x-polarized central spin to a combination of all possible state with appropriate coefficients:
\[U_{\rm c}\vert s\rangle^x_c=\sum_{l=0}^{2s}(-i)^l ~^{2s}C_l\vert s-l\rangle^x_c\]
Combining the effect of $U_{\rm sat}$ and $U_{\rm c}$ we get:\\
\begin{align}
    &U_d\vert \psi(0) \rangle=\vert \psi(T/2) \rangle\\ \nonumber &=\prod_{i=1}^{N_{\rm sat}}\left(\vert {\rm +x} \rangle_s^i-i\vert {\rm -x} \rangle_s^i\right)\otimes \sum_{l=0}^{2s}~^{2s}C_l(-i)^l\vert s-l\rangle^x_c \\ \nonumber 
    &=\sum_{k,l} (-i)^{k+l}\vert k \rangle_s\otimes ~^{2s}C_l\vert s-l\rangle^x_c
    \end{align}

The sum implies a sum over $k$ from 0 to $N_{\rm sat}$ and a sum over $l$ from 0 to $2s$. $\vert k \rangle_s$ denotes the combination of all states where $k$ out of $N_{\rm sat}$ spins are polarized in the $-x$ direction. $\vert s-l\rangle^{\alpha}_c$ is the $l+1^{th}$ eigenstate of $S_c^{\alpha}$.\\

From $t=T/2$ to $t=T$, as a result of $U_0$ each $\vert k \rangle_s \vert s-l\rangle^x_c$ state will accumulate a phase $-i\dfrac{\pi}{2}(N_{\rm sat}-2k)(s-l)$.\\

From magnetization plots, we can see that the system shows different kinds of DTC phases depending on whether $N_{\rm sat}$ is odd or even and $s$ is a full integer or a half-integer. Hence we will analyze these cases separately.\\

\textbf{Case I: When $N_{\rm sat}$ is even and $s$ is a full integer:}\\

$N_{\rm sat}=2m$: the phase factor becomes $-i\pi (m-k)(s-l)=\zeta_1=\pm1$ depending on whether the two factors are even or odd. Along with that, the $(-i)^{k+l}$ factor in eq 14 becomes $\pm1=\zeta_2$ ($\pm i =\zeta_2 i$) when both $j$ and $k$ are even or odd (when one of them is even and the other one is odd). Incorporating these factors, we can see that $\psi(T)$ will become:\\

\begin{align}
    &~^{2s}C_l\Bigg[\sum_{\substack{even~k\\even~l}}\zeta_1 \zeta_2 \vert k \rangle_s \vert s-l\rangle^x_c+\sum_{\substack{even~k\\odd~l}}\zeta_1 \zeta_2i \vert k \rangle_s \vert s-l\rangle^x_c\\ \nonumber &+\sum_{\substack{odd~k\\even~l}}\zeta_1 \zeta_2i \vert k \rangle_s \vert s-l\rangle^x_c+\sum_{\substack{odd~k\\odd~l}}\zeta_1 \zeta_2 \vert k \rangle_s \vert s-l\rangle^x_c \Bigg]\\ \nonumber
    &=\sum_{even~k}\zeta_1 \vert k \rangle_s ~^{2s}C_l \bigg[\sum_{even~l}\zeta_2 \vert s-l \rangle^x_c + \sum_{odd~l}\zeta_2 i \vert s-l \rangle^x_c\bigg]\\ \nonumber &+\sum_{odd~k}\zeta_1 i\vert k \rangle_s ~^{2s}C_l \bigg[\sum_{even~l}\zeta_2 \vert s-l \rangle^x_c - \sum_{odd~l}\zeta_2 i \vert s-l \rangle^x_c\bigg]
    \end{align}

For spin $s$, let the $S^{\alpha}$ eigenstates with the highest and the lowest eigenvalue be denoted by $\vert s\rangle^{\alpha}$ and $\vert -s\rangle^{\alpha}$. Then we can write these states using the complete basis made up of $S^x$ eigenstates:\\
\begin{align}
    \label{basis-transformation}
    &\vert \pm s\rangle^y = \sum_{l=0}^{2s}(\pm i)^l~^{2s}C_l\vert s-l\rangle^x \\ \nonumber
    &\vert \pm s\rangle^z = \sum_{l=0}^{2s}(\pm 1)^l~^{2s}C_l\vert s-l\rangle^x.
\end{align}

For spin half, this becomes $\vert \pm y\rangle=\vert x \rangle\pm i\vert -x \rangle$. Hence, for the satellite spins: $\prod_{j=1}^N\vert { \pm y} \rangle^j_s=\prod_{j=1}^N\bigg(\vert+x\rangle^j_s\pm i\vert -x\rangle^j_s\bigg)$. Using these expressions, $\psi(T)$ will become \\

\begin{align}
    &\bigg[\prod_{i=1}^{N_{\rm sat}}\vert {\rm +y} \rangle_s^i+\prod_{i=1}^{N_{\rm sat}}\vert {\rm -y} \rangle_s^i\bigg]\vert +s \rangle_c^y+\\ \nonumber &\bigg[\prod_{i=1}^{N_{\rm sat}}\vert {\rm +y} \rangle_s^i-\prod_{i=1}^{N_{\rm sat}}\vert {\rm -y} \rangle_s^i\bigg]\vert -s \rangle_c^y \\ \nonumber
    &=\left(\prod_{i=1}^{N_{\rm sat}}\vert {\rm +y} \rangle_s^i\right)\vert+s\rangle_c^y+\left(\prod_{i=1}^{N_{\rm sat}}\vert {\rm -y} \rangle_s^i\right)\vert +s \rangle_c^y\\ \nonumber &+\left(\prod_{i=1}^{N_{\rm sat}}\vert {\rm +y} \rangle_s^i\right)\vert -s \rangle_c^y-\left(\prod_{i=1}^{N_{\rm sat}}\vert {\rm -y} \rangle_s^i\right)\vert -s \rangle_c^y
\end{align}
Now, the application of the second kick, i.e. a rotation of $\pi/2$ about the $z$ axis will take all the $|\pm y\rangle$ states to $|\mp x\rangle$ and thus\\
\begin{align}
        |\psi(3T/2)\rangle=&\left(\prod_{i=1}^{N_{\rm sat}}\vert {\rm -x} \rangle^s_i\right) \vert -s\rangle^x_c+\left(\prod_{i=1}^{N_{\rm sat}}\vert {\rm +x} \rangle^s_i\right) \vert-s\rangle^x_s\\ \nonumber &+\left(\prod_{i=1}^{N_{\rm sat}}\vert {\rm -x} \rangle^s_i\right) \vert +s \rangle^x_s-\left(\prod_{i=1}^{N_{\rm sat}}\vert {\rm +x} \rangle^s_i\right) \vert +s \rangle^x_c
\end{align}
From $t=3T/2$ to $t=2T$ the first and the fourth terms will gain a phase factor of $\exp{(-i\pi\dfrac{s\times N_{sat}}{2})}$ and the second and the third terms will gain a phase factor of $\exp{(+i\pi\dfrac{s\times N_{sat}}{2})}$. Both of them are equal to 
$1$, hence \\
\begin{align}
        |\psi(2T)\rangle=&\left(\prod_{i=1}^{N_{\rm sat}}\vert {\rm +x} \rangle^s_i\right) \vert +s \rangle^x_c-\left(\prod_{i=1}^{N_{\rm sat}}\vert {\rm +x} \rangle^s_i\right) \vert -s \rangle^x_s\\ \nonumber &-\left(\prod_{i=1}^{N_{\rm sat}}\vert {\rm -x} \rangle^s_i\right) \vert +s\rangle^x_s+\left(\prod_{i=1}^{N_{\rm sat}}\vert {\rm -x} \rangle^s_i\right) \vert -s \rangle^x_c
\end{align}
To compute the state at $t=3T$ we will look the conversion of $|+x\rangle|+x\rangle$. By going through the same recipe it's possible to show that the other 3 states will go through the following conversion:\\
\begin{align}
    |+x\rangle|+x\rangle &\rightarrow |+y\rangle|+y\rangle+|+y\rangle|-y\rangle\\ \nonumber &+|-y\rangle|+y\rangle-|-y\rangle|-y\rangle \\ \nonumber
    -|-x\rangle|-x\rangle &\rightarrow -|-y\rangle|-y\rangle-|+y\rangle|-y\rangle\\ \nonumber &-|-y\rangle|+y\rangle+|+y\rangle|+y\rangle \\ \nonumber
    -|+x\rangle|-x\rangle &\rightarrow -|+y\rangle|-y\rangle-|-y\rangle|-y\rangle\\ \nonumber &-|+y\rangle|+y\rangle+|-y\rangle|+y\rangle \\ \nonumber
    -|-x\rangle|+x\rangle &\rightarrow -|-y\rangle|+y\rangle-|+y\rangle|+y\rangle\\ \nonumber &-|-y\rangle|-y\rangle+|+y\rangle|-y\rangle     
\end{align}
Sum of these states will give us $\psi(3T)=\left(\prod_{i=1}^{N_{\rm sat}}\vert {\rm -y} \rangle^s_i\right) \vert { -s} \rangle^y_c$. The fourth kick will take $|-y\rangle$ states to $\vert +x\rangle$. Hence after 4 periods, the system will return to its initial x-polarized state, thus exhibiting a 4-period DTC phase.\\

\textbf{Case II: When $N_{\rm sat}$ is odd and $s$ is a full integer:}\\

$N_{\rm sat}=2m+1$, $s$ is a full integer: the phase factor becomes $\dfrac{-i\pi}{2} (2m-2k-1)(s-l)=\dfrac{-i\pi}{2}(s-l)\times{\rm an ~odd ~number}$. This becomes $\pm i=\zeta_1i~(\pm1=\zeta_11)$ when $(s-l)$ is odd (even). The factor $(-i)^{k+l}$ remains the same: $\pm1=\zeta_21$ ($\pm i =\zeta_2 i$) when both $j$ and $k$ are even or odd (when one of them is even and the other is odd). Incorporating these factors, we can see that $\psi(T)$ will become:\\

\begin{align}
    &~^{2s}C_l \Bigg[\sum_{\substack{even~k\\even~l}}\zeta_1 \zeta_2 \vert k \rangle_s \vert s-l\rangle^x_c+\sum_{\substack{even~k\\odd~l}}\zeta_1 \zeta_2 \vert k \rangle_s \vert s-l\rangle^x_c\\ \nonumber &+\sum_{\substack{odd~k\\even~l}}\zeta_1 \zeta_2i \vert k \rangle_s \vert s-l\rangle^x_c+\sum_{\substack{odd~k\\odd~l}}\zeta_1 \zeta_2i \vert k \rangle_s \vert s-l\rangle^x_c\Bigg]\\ \nonumber
    &=\bigg[\sum_{even~k}\zeta_2 \vert k \rangle_s + \sum_{odd~k}\zeta_2 i \vert k \rangle_s\bigg]~^{2s}C_l\sum_{even~l}\zeta_1 \vert s-l \rangle^z_c\\ \nonumber &+\bigg[\sum_{even~k}\zeta_2 \vert k \rangle_s - \sum_{odd~k}\zeta_2 i \vert k \rangle_s\bigg]~^{2s}C_l\sum_{odd~l}\zeta_1 \vert s-l \rangle^z_c\\ \nonumber
    &=\prod_{i=1}^{N_{\rm sat}}\vert {\rm +y} \rangle_s^i\bigg[\vert+s \rangle_c^z+\vert -s \rangle_c^z\bigg]-\prod_{i=1}^{N_{\rm sat}}\vert {\rm -y} \rangle_s^i\bigg[\vert+s\rangle_c^z-\vert-s \rangle_c^z\bigg]
\end{align} 
The last equality can be understood using eq \eqref{basis-transformation}. Hence, the system becomes a combination of 4 states:\\
\begin{align}
        |\psi(T)\rangle=&\left(\prod_{i=1}^{N_{\rm sat}}\vert {\rm +y} \rangle_s^i\right) \vert+s \rangle_c^z+\left(\prod_{i=1}^{N_{\rm sat}}\vert {\rm +y} \rangle_s^i\right) \vert-s \rangle_c^z\\ \nonumber &-\left(\prod_{i=1}^{N_{\rm sat}}\vert {\rm -y} \rangle_s^i\right)\vert+s \rangle_c^z+\left(\prod_{i=1}^{N_{\rm sat}}\vert {\rm -y} \rangle_s^i\right) \vert-s \rangle_c^z
\end{align}
The second kick from time $T$ to $3T/2$ will take the $\vert \pm y \rangle $ states to $\vert \mp x \rangle$ states and keep the $\vert \pm z\rangle$ states along the z direction, as mentioned earlier. Now, if we follow the same recipe to track the phase factors, at time $2T$ the system will end up at:\\
\begin{align}
    \vert \psi(2T) \rangle&=\left(\prod_{i=1}^{N_{\rm sat}}\vert {\rm +x} \rangle_s^i\right)\vert+s \rangle_c^y+\left(\prod_{i=1}^{N_{\rm sat}}\vert {\rm +x} \rangle_s^i\right)\vert-s \rangle_c^y\\ \nonumber &-i\left(\prod_{i=1}^{N_{\rm sat}}\vert {\rm -x} \rangle_s^i\right)\vert+s \rangle_c^y+i\left(\prod_{i=1}^{N_{\rm sat}}\vert {\rm -x} \rangle_s^i\right) \vert-s \rangle_c^y
\end{align}
Similarly after the $3^{rd}$ period the system will reach\\
\begin{align}
    \vert \psi(3T) \rangle
    &=\left(\prod_{i=1}^{N_{\rm sat}}\vert {\rm +y} \rangle_s^i\right) \vert+s \rangle_c^x-\left(\prod_{i=1}^{N_{\rm sat}}\vert {\rm +y} \rangle_s^i\right) \vert-s \rangle_c^x\\ \nonumber &-i\left(\prod_{i=1}^{N_{\rm sat}}\vert {\rm -y} \rangle_s^i\right) \vert-s \rangle_c^x-i\left(\prod_{i=1}^{N_{\rm sat}}\vert {\rm -y} \rangle_s^i\right) \vert-s \rangle_c^x
\end{align}
And after the $4^{th}$ kick
\begin{align}
    \vert \psi(4T) \rangle
    &=\left(\prod_{i=1}^{N_{\rm sat}}\vert {\rm +x} \rangle_s^i\right) \vert+s \rangle_c^z-\left(\prod_{i=1}^{N_{\rm sat}}\vert {\rm +x} \rangle_s^i\right) \vert-s \rangle_c^z\\ \nonumber &-i\left(\prod_{i=1}^{N_{\rm sat}}\vert {\rm -x} \rangle_s^i\right) \vert+s \rangle_c^z-i\left(\prod_{i=1}^{N_{\rm sat}}\vert {\rm -x} \rangle_s^i\right) \vert-s \rangle_c^z
\end{align}
Interestingly, following the recipe one more time will polarize the state in the y direction: $\vert \psi(5T)\rangle=\left(\prod_{i=1}^{N_{\rm sat}}\vert {\rm +y} \rangle_s^i\right) \vert+s \rangle_c^y$. As mentioned before, another application of $U_d$ will take the y polarized spins to $-x$ direction. Hence $\vert \psi(11T/2)\rangle=\left(\prod_{i=1}^{N_{\rm sat}}\vert {\rm -x} \rangle_s^i\right) \vert-s \rangle_c^x$. Applying $U_0$ will just add a phase factor to the state. Thus, after $6T$ the system will reverse its polarization and mark the half period. After 6 more periods, the system will reverse its polarization again to go back to its initial state. Hence we can show that the system shows a 12-period DTC phase.\\

\textbf{Case III: When $N_{\rm sat}$ is even and $s$ is a half integer:}\\

$N_{\rm sat}=2m$, $s$ is a half integer: the phase factor becomes $\dfrac{-i\pi}{2}(2m-2k)(s-l)=(\dfrac{-i\pi}{2})2\times(s-l)(m-k)=(\dfrac{-i\pi}{2})(m-k)\times {\rm~an~odd~number}$ which is exactly similar to the previous case except that the odd number in the phase factor comes from the central spin instead of the satellite spins. The dynamics in this case will be exactly like the previous case, i.e. the system will still show a 12-period DTC phase except that the polarization of the central spin and the satellite spins will be interchanged. So, if we start with all $+x$ polarized spins like in the previous cases, $\vert \psi(T)\rangle$ will become 
\begin{align}
        |\psi(T)\rangle=&\left(\prod_{i=1}^{N_{\rm sat}}\vert {\rm +z} \rangle_s^i\right) \vert+s \rangle_c^y+\left(\prod_{i=1}^{N_{\rm sat}}\vert {\rm +z} \rangle_s^i\right) \vert-s \rangle_c^y\\ \nonumber &-\left(\prod_{i=1}^{N_{\rm sat}}\vert {\rm -z} \rangle_s^i\right) \vert+s \rangle_c^y+\left(\prod_{i=1}^{N_{\rm sat}}\vert {\rm -z} \rangle_s^i\right) \vert-s \rangle_c^y
\end{align}
Following the trend, at $t=2T$ the satellite spins will be polarized along the y axis and the central spin will be polarized along the x axis. We will see the same thing at $t=3T$ and $4T$. At $t=5T$ both the central and the satellite spins will be polarized in the $y$ direction. And just like the previous case, the $6^{th}$ period will take all the spins to $-x$ direction to indicate the completion of half period.\\

\textbf{Case IV: When $N_{\rm sat}$ is odd and $s$ is a half integer:}\\

$N_{\rm sat}=2m+1$: the phase factor becomes $-i\dfrac{\pi}{2} (2m-2k-1)(s-l)=i\dfrac{\pi}{4} (2m-2k-1)(2s-2l)=-i\dfrac{\pi}{4}\times pq$ where $p$ and $q$ are both odd numbers. So, the phase factor will be $\pm i\pm1=\zeta_1i+\zeta_2$. And of course, $(-i)^{k+l}=\zeta_3 i$ or $\zeta_3$ depending on the parity of $k$ and $l$. In this case $\vert \psi(T)\rangle$ will be\\
 \begin{align}
    &~^{2s}C_l \Bigg[\sum_{\substack{even~k\\even~l}}(\zeta_1i+\zeta_2)\zeta_3 \vert k \rangle_s \vert s-l\rangle_c^x+\sum_{\substack{even~k\\odd~l}}(\zeta_1i+\zeta_2)\zeta_3i \vert k \rangle_s \vert s-l\rangle_c^x\\ \nonumber &
    +\sum_{\substack{odd~k\\even~l}}(\zeta_1i+\zeta_2)\zeta_3i \vert k \rangle_s \vert s-l\rangle_c^x+\sum_{\substack{odd~k\\odd~l}}(\zeta_1i+\zeta_2)\zeta_3 \vert k \rangle_s \vert s-l\rangle_c^x\Bigg]\\ \nonumber
    &=~^{2s}C_l \Bigg[\sum_{\substack{even~k\\even~l}}-(1+i) \vert k \rangle_s \vert s-l\rangle_c^x+\sum_{\substack{even~k\\odd~l}}-(1+i) \vert k \rangle_s \vert s-l\rangle_c^x\\ \nonumber &+\sum_{\substack{odd~k\\even~l}}(1+i) \vert k \rangle_s \vert s-l\rangle_c^x+\sum_{\substack{odd~k\\odd~l}}-(1+i)\vert k \rangle_s \vert s-l\rangle_c^x\Bigg]\\ \nonumber
    &=\bigg[\sum_{even~k}\vert k \rangle_s + \sum_{odd~k} \vert k \rangle_s\bigg]~^{2s}C_l\sum_{odd~l} -(1+i)\vert s-l \rangle_c^x\\ \nonumber &+\bigg[\sum_{even~k}\vert k \rangle_s - \sum_{odd~k} \vert k \rangle_s\bigg]~^{2s}C_l\sum_{even~l}-(1+i) \vert s-l \rangle_c^x\\ \nonumber
    &=(1+i)\prod_{i=1}^{N_{\rm sat}}\vert {\rm +z} \rangle_s^i\bigg[\vert +s \rangle_c^z-\vert -s \rangle_c^z\bigg]\\ \nonumber &-(1+i)\prod_{i=1}^{N_{\rm sat}}\vert {\rm -z} \rangle_s^i\bigg[\vert +s \rangle_c^z+\vert -s \rangle_c^z\bigg]
\end{align} 
\begin{align}
        |\psi(T)\rangle=&(1+i)\left(\prod_{i=1}^{N_{\rm sat}}\vert {\rm +z} \rangle_s^i\right) \vert +s \rangle_c^z-(1+i)\left(\prod_{i=1}^{N_{\rm sat}}\vert {\rm +z} \rangle_s^i\right)\vert -s \rangle_c^z\\ \nonumber &-(1+i)\left(\prod_{i=1}^{N_{\rm sat}}\vert {\rm -z} \rangle_s^i\right) \vert +s \rangle_c^z-(1+i)\left(\prod_{i=1}^{N_{\rm sat}}\vert {\rm -z} \rangle_s^i\right) \vert -s \rangle_c^z
\end{align}
Here, the second equality can be understood if we notice that some of the terms from the top-most expression will get cancelled. If we follow the procedure outlined in the previous cases, after the second kick we will see:\\
\begin{align}
        |\psi(2T)\rangle=&\left(\prod_{i=1}^{N_{\rm sat}}\vert {\rm +y} \rangle_s^i\right) \vert +s \rangle_c^y-\left(\prod_{i=1}^{N_{\rm sat}}\vert {\rm +y} \rangle_s^i\right) \vert -s \rangle_c^y\\ \nonumber &-\left(\prod_{i=1}^{N_{\rm sat}}\vert {\rm -y} \rangle_s^i\right) \vert  +s \rangle_c^y-\left(\prod_{i=1}^{N_{\rm sat}}\vert {\rm -y} \rangle_s^i\right) \vert  -s \rangle_c^y
\end{align}
Applying $U_d$ on this means all the $\vert \pm y\rangle$ states will become $\vert \mp x\rangle$ and as a result of a further $U_0$ each term will accumulate a phase factor $\pm i\dfrac{\pi}{2}sN_{\rm sat}=\pm1\pm  i$:\\
\begin{align}
        |\psi(3T)\rangle=&-(1-i)\left(\prod_{i=1}^{N_{\rm sat}}\vert {\rm +x} \rangle_s^i\right)\vert +s \rangle_c^x\\ \nonumber &-(1+i)\left(\prod_{i=1}^{N_{\rm sat}}\vert {\rm +x} \rangle_s^i\right)\vert -s \rangle_c^x\\ \nonumber &+(1+i)\left(\prod_{i=1}^{N_{\rm sat}}\vert {\rm -x} \rangle_s^i\right)\vert +s \rangle_c^x\\ \nonumber &-(1-i)\left(\prod_{i=1}^{N_{\rm sat}}\vert {\rm -x} \rangle_s^i\right) \vert -s \rangle_c^x
\end{align}
To compute the state at $t=4T$ we will look at the evolution of $|+x\rangle|+x\rangle$ in eq 26. By going through the same recipe it's possible to show that the states in eq 29 will go through the following conversion:\\
\begin{align}
    |+x\rangle|+x\rangle &\rightarrow (1+i)|+z\rangle|+z\rangle-(1+i)|-z\rangle|+z\rangle\\ \nonumber &-(1+i)|+z\rangle|-z\rangle-(1+i)|-z\rangle|-z\rangle \\ \nonumber
    |-x\rangle|+x\rangle &\rightarrow   (1-i)|+z\rangle|+z\rangle-(1-i)|-z\rangle|+z\rangle\\ \nonumber &+(1-i)|+z\rangle|-z\rangle+(1-i)|-z\rangle|-z\rangle \\ \nonumber
    |+x\rangle|-x\rangle &\rightarrow   -(1-i)|+z\rangle|+z\rangle-(1-i)|-z\rangle|+z\rangle\\ \nonumber &+(1-i)|+z\rangle|-z\rangle-(1-i)|-z\rangle|-z\rangle \\ \nonumber
    |-x\rangle|-x\rangle &\rightarrow   -(1+i)|+z\rangle|+z\rangle-(1+i)|-z\rangle|+z\rangle\\ \nonumber &-(1+i)|+z\rangle|-z\rangle+(1+i)|-z\rangle|-z\rangle \\ \nonumber
\end{align}
The sum of these states with appropriate coefficients will give us:
\begin{align}
        |\psi(4T)\rangle=&-\left(\prod_{i=1}^{N_{\rm sat}}\vert {\rm +z} \rangle_s^i\right)\vert +s \rangle_c^z+\left(\prod_{i=1}^{N_{\rm sat}}\vert {\rm +z} \rangle_s^i\right)\vert -s \rangle_c^z\\ \nonumber &+\left(\prod_{i=1}^{N_{\rm sat}}\vert {\rm -z} \rangle_s^i\right)\vert +s \rangle_c^z-\left(\prod_{i=1}^{N_{\rm sat}}\vert {\rm -z} \rangle_s^i\right)\vert -s \rangle_c^z\\ \nonumber
        &=\Bigg[\left(\prod_{i=1}^{N_{\rm sat}}\vert {\rm -z} \rangle_s^i\right)-\left(\prod_{i=1}^{N_{\rm sat}}\vert {\rm +z} \rangle_s^i\right)\Bigg]\\ \nonumber &\otimes\Bigg[\vert +s \rangle_c^z-\vert -s \rangle_c^z\Bigg]
\end{align}
This is a much more symmetric state than $\vert \psi(T)\rangle$ and the application of $U_dU_0$ will produce a cat state in the y direction:\\
\begin{align}
    \vert \psi(5T)\rangle=i\left(\prod_{i=1}^{N_{\rm sat}}\vert {\rm +y} \rangle_s^i\right)\vert +s \rangle_c^y+\left(\prod_{i=1}^{N_{\rm sat}}\vert {\rm -y} \rangle_s^i\right)\vert -s \rangle_c^y
\end{align}
The $6^{th}$ $U_d$ will take $\vert \pm y\rangle$ states to $\vert \mp x\rangle$ and the subsequent $U_0$ will again just add a phase factor $\pm i\dfrac{\pi}{2}sN_{\rm sat}=\pm1\pm  i$:\\
\begin{align}
    \label{psi_5T}
    \vert \psi(6T)\rangle&=(-1-i)\left(\prod_{i=1}^{N_{\rm sat}}\vert {\rm +x} \rangle_s^i\right)\vert +s \rangle_c^x\\ \nonumber &+(1-i)\left(\prod_{i=1}^{N_{\rm sat}}\vert {\rm -x} \rangle_s^i\right) \vert -s \rangle_c^x
\end{align}
This marks the completion of the quarter period. eq \ref{psi_5T} can be written as:
\begin{align}
    U^6\left(\prod_{i=1}^{N_{\rm sat}}\vert {\rm +x} \rangle_s^i\right) \vert +s \rangle_c^x&=(-1-i)\left(\prod_{i=1}^{N_{\rm sat}}\vert {\rm +x} \rangle_s^i\right) \vert +s \rangle_c^x\\ \nonumber &+(1-i)\left(\prod_{i=1}^{N_{\rm sat}}\vert {\rm -x} \rangle_s^i\right)\vert -s \rangle_c^x
\end{align}
Following similar steps, we can show\\
\begin{align}
    \label{psi_6T_mm}
    U^6\left(\prod_{i=1}^{N_{\rm sat}}\vert {\rm -x} \rangle_s^i\right)\vert -s \rangle_c^x&=(-1-i)\left(\prod_{i=1}^{N_{\rm sat}}\vert {\rm -x} \rangle_s^i\right) \vert -s \rangle_c^x\\ \nonumber &+(1-i)\left(\prod_{i=1}^{N_{\rm sat}}\vert {\rm +x} \rangle_s^i\right)\vert +s \rangle_c^x
\end{align}

\begin{align}
    \label{psi_12T}
    \vert \psi(12T)\rangle= U^6\vert \psi(6T)\rangle&=(-1-i)U^6\left(\prod_{i=1}^{N_{\rm sat}}\vert {\rm +x} \rangle_s^i\right)\vert +s \rangle_c^x\\ \nonumber &+(1-i)U^6\left(\prod_{i=1}^{N_{\rm sat}}\vert {\rm -x} \rangle_s^i\right)\vert -s \rangle_c^x
\end{align}
Using \eqref{psi_6T_mm} and \eqref{psi_12T}: $\vert \psi(12T)\rangle=\left(\prod_{i=1}^{N_{\rm sat}}\vert {\rm -x} \rangle_s^i\right)\vert -s \rangle_c^x$. In this case, too, polarization reversal marks half period, but we can see that it takes 12 periods. Another 12 periods will take the system back to its original state, thus exhibiting a 24-period DTC phase.\\

\section*{OTHER HO-DTC PHASES}

As mentioned in section IV, there are four points in the $\lambda \sim g$ phase diagram where special higher-order DTC phase can be seen. Around each of those points there are 4 symmetrically placed points where we can see two more types of HO-DTC phase. In these phases the system does not go through a set of highly symmetric super-cat states like we have seen in the case of the special HO-DTC phase. Hence, we call them regular HO-DTC phases. There are two different types of regular HO-DTC phases. Their time-periods and conditions regarding the value of $N_{\rm sat}$ and $s$ are given in tables \ref{tab_ho-dtc2} \ref{tab_ho-dtc3}. These states can be seen at specific points in the phase diagram. The list is given in table \ref{tab_ho-dtc_reg}. The magnetization and entanglement entropy variation for these phases are shown in fig \ref{ho-dtc2} and \ref{ho-dtc3}.\\

\begin{table}[]
    \centering
    \begin{tabular}{|c|c|c|}\hline
         Special DTC point & Regular DTC points & Regular DTC class  \\ \hline\hline
        \multirow{2}{*}{$(\pi,\pi/2)$} & \multicolumn{1}{c|}{$(\pi,\pi/4$), ($\pi,3\pi/4$)} & 1 \\\cline{2-3}
                                     & \multicolumn{1}{c|}{$\pi/2,\pi/2$, $3\pi/2,\pi/2$} & 2 \\   \hline
        \multirow{2}{*}{$(3\pi,\pi/2)$} & \multicolumn{1}{c|}{$3\pi,\pi/4$, $3\pi,3\pi/4$} & 1 \\\cline{2-3}
                                     & \multicolumn{1}{c|}{$5\pi/2,\pi/2$, $7\pi/2,\pi/2$} & 2 \\   \hline 
        \multirow{2}{*}{$(\pi,3\pi/2)$} & \multicolumn{1}{c|}{$(\pi,5\pi/4$), ($\pi,7\pi/4$)} & 1 \\\cline{2-3}
                                     & \multicolumn{1}{c|}{$\pi/2,3\pi/2$, $3\pi/2,3\pi/2$} & 2 \\   \hline
        \multirow{2}{*}{$(3\pi,3\pi/2)$} & \multicolumn{1}{c|}{$3\pi,5\pi/4$, $3\pi,7\pi/4$} & 1 \\\cline{2-3}
                                     & \multicolumn{1}{c|}{$5\pi/2,3\pi/2$, $7\pi/2,3\pi/2$} & 2 \\   \hline                             \end{tabular}
    \caption{regular HO-DTC phase points in the $\lambda \sim g$ phase diagram}
    \label{tab_ho-dtc_reg}
\end{table}
\begin{table}[]
    \centering
    \begin{tabular}{|c|c|c|}\hline
         $N_{\rm sat}$ & $s$ & DTC period  \\ \hline\hline
        $4j$ & even integer & 24\\
        $4j$ & odd integer & 24\\
        $4j+2$ & even integer & 24\\
        $4j+2$ & odd integer & 12\\ \hline
    \end{tabular}
    \caption{The dependence of the time-period of the DTC on $N_{\rm sat}$ and $s$ for the class I regular HO-DTC}
    \label{tab_ho-dtc2}
\end{table}
\begin{figure*}
    \centering
    \includegraphics[width=0.9\textwidth]{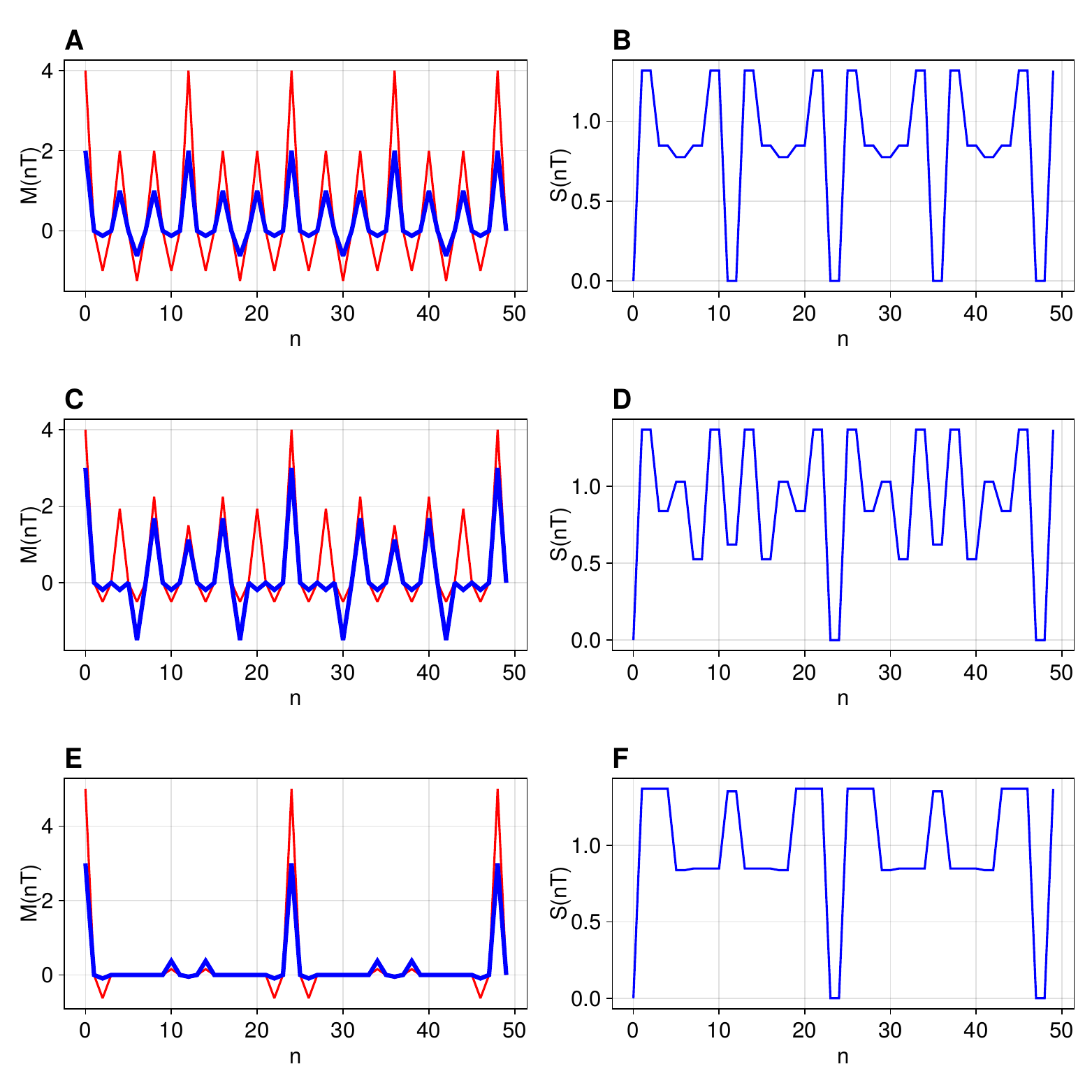}
    \caption{Example of class I regular HO-DTC: Magnetization evolution with time for $\lambda=\pi/2,~g=\pi/2$. Top row: $N_{\rm sat}=8;~s=4$, Middle row: $N_{\rm sat}=8;~s=6$, Bottom row: $N_{\rm sat}=10;~s=4$. The left panel shows the magnetization, and the right panel shows the entanglement entropy variation. The blue line denotes central spin magnetization, and the red line denotes satellite spin magnetization.}
    \label{ho-dtc2}
\end{figure*}

\begin{table}[]
    \centering
    \begin{tabular}{|c|c|c|}\hline
         $N_{\rm sat}$ & $s$ & DTC period  \\ \hline\hline
        $4j$ & even integer & 12\\
        $4j$ & odd integer & 24\\
        $4j+2$ & even integer & 24\\
        $4j+2$ & odd integer & 24\\ \hline
    \end{tabular}
    \caption{The variation of time period depending on $N_{\rm sat}$ and $s$ for the class II regular HO-DTC}
    \label{tab_ho-dtc3}
\end{table}

\begin{figure*}
    \centering
    \includegraphics[width=0.9\textwidth]{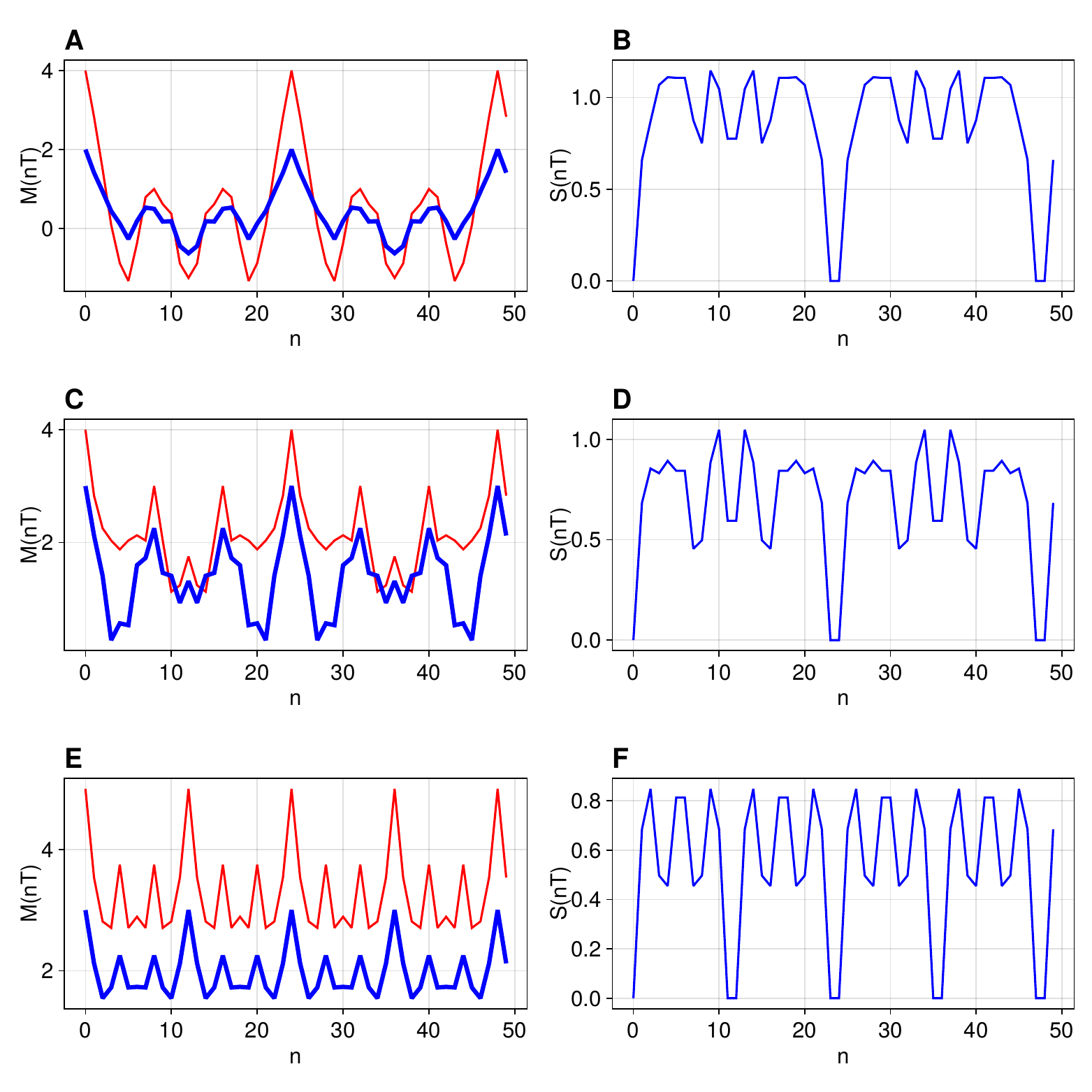}
    \caption{Example of class II regular HO-DTC: Magnetization evolution with time for $\lambda=\pi,~g=\pi/4$.  Top row: $N_{\rm sat}=8;~s=4$, Middle row: $N_{\rm sat}=8;~s=6$, Bottom row: $N_{\rm sat}=10;~s=6$. The left panel shows the magnetization, and the right panel shows the entanglement entropy variation. The blue line denotes central spin magnetization, and the red line denotes satellite spin magnetization.}
    \label{ho-dtc3}
\end{figure*}

\end{widetext}

\end{document}